\documentclass[journal]{IEEEtran}

\usepackage{url}
\usepackage[cmex10]{amsmath}
\usepackage[pdftex]{graphicx}
\usepackage{graphicx,epsfig}
\usepackage{epstopdf}
\usepackage{cite}
\usepackage{stfloats}
\usepackage{color}
\usepackage{amsfonts,helvet}

\usepackage[tight,footnotesize]{subfigure}
\usepackage[usenames,dvipsnames,svgnames,table]{xcolor}

\usepackage{balance}

\newtheorem{theorem}{Theorem}

\newtheorem{algo}[theorem]{Algorithm}

\newtheorem{proposition}[theorem]{Proposition}

\newcounter{proposition}
\begin{document}

\title{QoE-aware Scalable Video Transmission in MIMO~Systems}

\author{Soo-Jin~Kim,~\IEEEmembership{Student Member,~IEEE,}
	Gee-Yong~Suk,~\IEEEmembership{Student Member,~IEEE,}\\
	Jong-Seok~Lee,~\IEEEmembership{Senior Member,~IEEE,}
	and Chan-Byoung~Chae,~\IEEEmembership{Senior Member,~IEEE}
	\thanks{S. Kim, G.-Y. Suk, J.-S. Lee, and C.-B. Chae are with School of Integrated Technology, Yonsei University, Korea (e-mail:\{soojin.kim, gysuk, jong-seok.lee, cbchae\}@yonsei.ac.kr).}}

\maketitle

\begin{abstract}
An important concept in wireless systems has been quality of experience (QoE)-aware video transmission. Such communications are considered not only connection-based communications but also content-aware communications, since the video quality is closely related to the content itself. It becomes necessary therefore for video communications to utilize a cross-layer design (also known as joint source and channel coding). To provide efficient methods of allocating network resources, the wireless network uses its cross-layer knowledge to perform unequal error protection (UEP) solutions. In this article, we summarize the latest video transmission technologies that are based on scalable video coding (SVC) over multiple-input multiple-output (MIMO) systems with cross-layer designs. To provide insight into video transmission in wireless networks, we investigate UEP solutions in the delivering of video over massive MIMO systems. Our results show that in terms of quality of experience (QoE), SVC layer prioritization, which was considered important in the prior work, is not always beneficial in massive MIMO systems; consideration must be given to the content characteristics. 
%
%
\end{abstract}

\begin{IEEEkeywords}
MIMO, massive MIMO, SVC, QoS, QoE,  perceptual quality, video transmission, resource allocation. 
\end{IEEEkeywords}

\section{Introduction}

With smart mobile devices dominating the world, wireless data traffic has increased exponentially. Keeping pace with such growth has been video traffic consumption, such as video on demand (VoD), IPTV, video call, video streaming, video sharing, and so on. An important research topic then is how best to maximize users' satisfaction with delivered video content.


In the past, physical and application layer technologies have been considered as separate dimensions. Physical layer technologies for 4G and 5G have been developed mostly as a way of increasing capacity; application layer technologies for video coding has been developed as a way to increase coding efficiency. This is not, however, the optimal way to increase efficiency in terms of the quality of transmitted video over wireless communications.

The degradation of video quality is caused by two major factors\textendash coding artifacts and network artifacts. Coding artifacts are caused by the application of lossy video data compression, which include blurring, blockiness, and ringing artifacts. Network artifacts are caused by packet losses preventing packets of video data from reaching their destinations. Such losses are caused by packet jitters, delays, drops, and so forth. These two types of artifacts are simultaneously involved in determining end user video quality. Therefore, it is important to first understand their combinational impact and then develop an optimal way of allocating resources to maximize users' satisfaction in video transmission. A natural choice to enhance the overall efficiency of video transmission in wireless networks is a cross-layer design.

In the physical layer, one way of expanding bandwidth and reducing error rate is with multiple-input multiple-output (MIMO) along with the aid of techniques known as spatial multiplexing and diversity~\cite{cb2007sigmag}. A powerful video compression technique that is well associated with MIMO systems is scalable video coding (SVC). In the application layer, SVC organizes the video data into a layered structure, enabling extraction of multiple video streams having different bitrates and levels of quality. For example, the base layer presents base quality; enhancement layers\textendash relying on the base layer's quality\textendash provide higher quality. The adaptability and scalability of SVC allow the selection of desirable target quality within the network constraints. 
Based on these two technologies, many studies have tried to develop video transmission frameworks that can find the optimal balance between efficiency and quality. In the prior work, cross-layer-designed video transmission frameworks, which are described in Section III, often adopted unequal error protection (UEP) solutions to efficiently allocate the resources. 
Popularly used UEP techniques are summarized below.

\begin{figure*}[!htbp]
\centering
\includegraphics[width=0.85\linewidth]{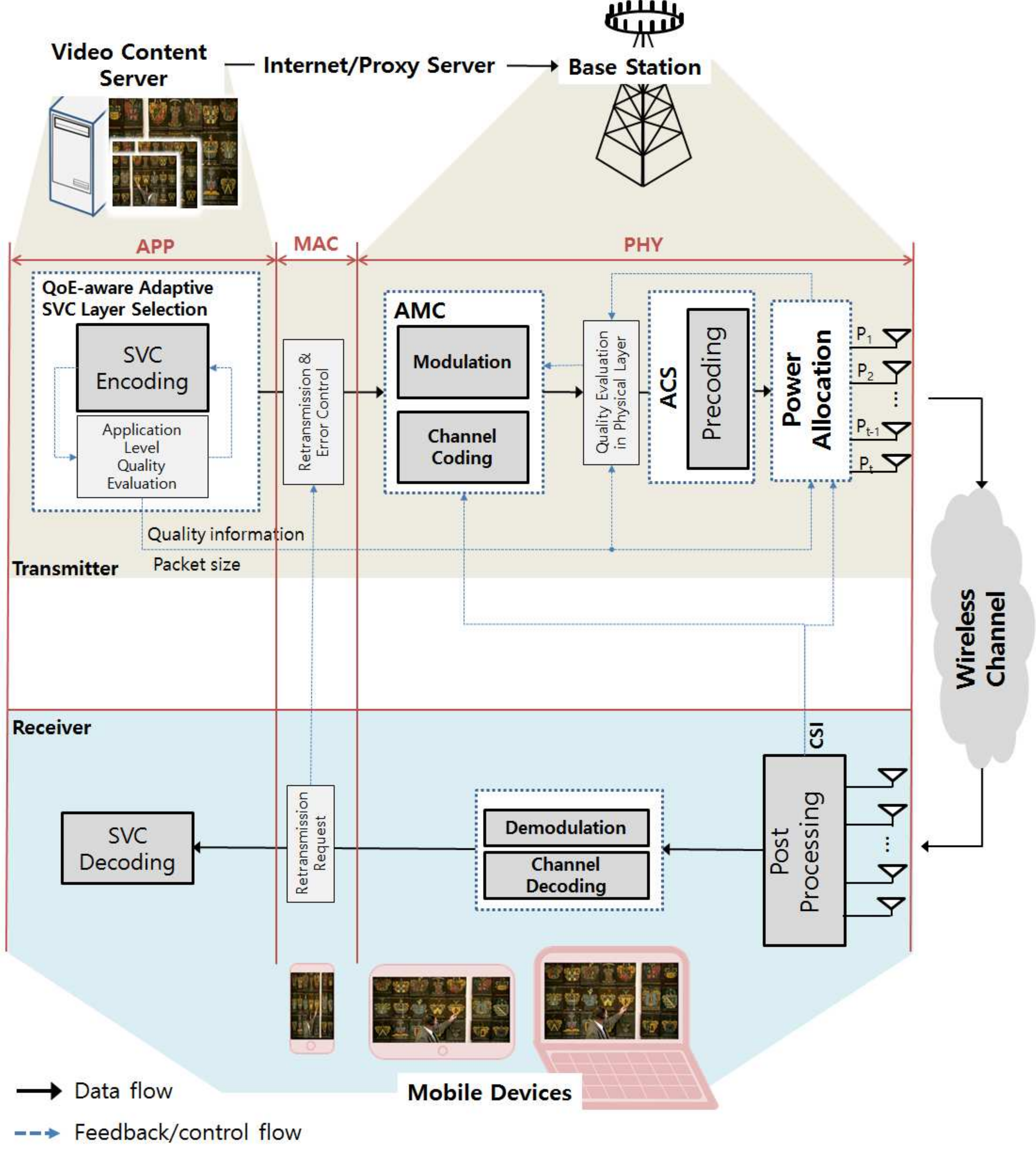}
   \caption{Overview of a QoE-aware video transmission system based on the cross-layer design.
      \label{fig1}}
 \end{figure*}

\begin{itemize}
\item \emph{Adaptive modulation and coding (AMC)}  
controls modulation order and coding rate per time slot, sub-block, or sub-channel.
The modulation order of a digital communication scheme is determined by the number of different symbols that can be transmitted using it. A higher modulation order refers to a higher throughput though it is more vulnerable to error than a lower modulation order. The code rate or error correction code (ECC) normally refers to the proportion of the data stream that is not redundant. A higher code rate means a higher throughput but a lower error correction. 
\item \emph{Adaptive channel selection (ACS)} assigns sub-channels for certain data transmissions. A certain SVC quality layer is often assigned to a specific sub-channel of the user. The ACS technique makes it easy to give priority to a SVC layer. 
\item \emph{Power allocation} assigns different transmit powers to the channels, sub-channels, or users.
Assigning a larger power to a channel or user results in a lower error rate or higher throughput of the channel or user. 
\end{itemize}

Figure 1 illustrates a QoE-aware video transmission system based on the cross-layer design.
In the figure, blocks for the aforementioned UEP techniques are shown on the transmitter side. In addition, blocks for quality evaluation are located in the application and physical layers to predict the video quality before sending. These can, however, be located at any place between the physical and application layers depending on the purpose of the video transmission framework.\\

In this article, we deal with the issue regarding how to optimize quality of experience (QoE) of end users for SVC transmission over massive MIMO systems. In particular, our contributions are summarized as follows:

\begin{itemize}
{\item We provide a broad discussion about video transmission algorithms for MIMO systems based on cross-layer design. This will enable our readers to understand current trends in this field.

\item Unlike prior work, this work conducts SVC transmission in massive MIMO systems, which is unique and the first of its kind in the literature.

\item We investigate the error characteristics of massive MIMO systems under a Rayleigh fading channel, which turned out to be different from that of MIMO systems. In Section IV, we derive the bit error probability formulation in massive MIMO systems with a zero-forcing (ZF) precoder. This allows the estimation of the error rate prior to transmission without any additional feedback; this saves resources that would otherwise have been used for limited feedback.

\item We show that different approaches should be considered when applying UEP techniques to massive MIMO systems. Unlike conventional MIMO systems, in massive MIMO systems, the priority exists not only at the base layer but also at other layers. With our simulation of unequal power allocation in massive MIMO systems, we show that to take full advantage of UEP for improved video quality at the user side, engineers must consider ``unconventional" power allocation mechanism between the layers.

\item Additionally, we introduce regression analysis as a possible approach to better clarify the relationship between video quality and assigned transmit power per SVC layer with content information. We believe this would serve as a basic guideline for future research. 
}
\end{itemize}

The rest of the article is organized as follows. Section~\ref{basic} describes the basic concepts of MIMO, SVC and error control methods in cross-layer design. Section~\ref{MIMO} explains video transmission techniques in MIMO systems. Section~\ref{massiveMIMO} presents our error rate analysis and a power allocation scheme in massive MIMO systems. Finally, we conclude the article in Section~\ref{Con}.


\begin{figure*}[t]
\centering
  \includegraphics[width=1\linewidth]{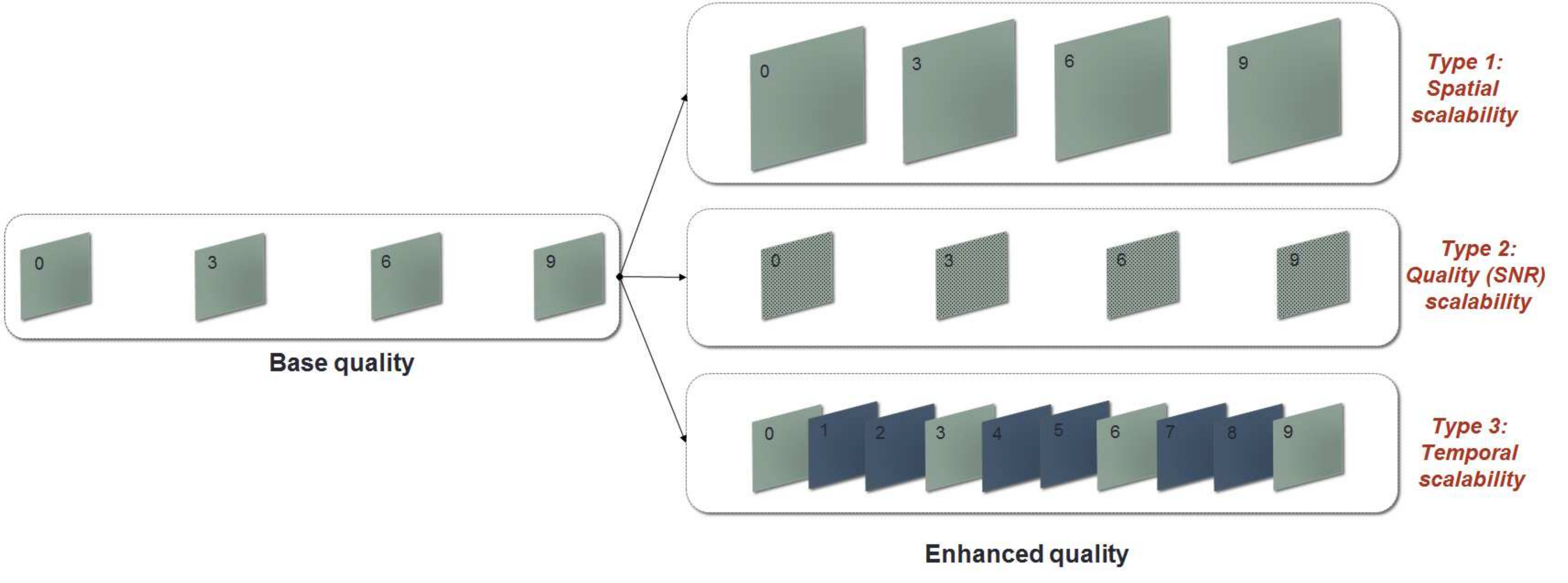}
   \caption{Three representative types of scalability.}
   \label{fig:SVC}
 \end{figure*}

\section{Basic concepts}
\label{basic}
Depending on the system optimization approaches, different methods can be selected for cross-layer design. This section explains the background of multiple antenna techniques and briefly goes over error-control methods and QoE-related knowledge of SVC. 
\subsection{MIMO}

In MIMO systems, the base station (BS) transmits a message signal to one or more users by using multiple antennas~\cite{Marzetta14}. It is important to suppress signal interference among channels. Precoding techniques are popularly used for this. Here, availability of the channel state information at the transmitter (CSIT) is considered importantly, particularly in video transmission, because a key to predicting the amount of video quality degradation is prior knowledge of network parameters, making it possible to utilize UEP solutions. Thus, perfect or partial availability of CSIT is usually assumed for video transmission in MIMO systems. 

A popular method of precoding based on CSIT is singular value decomposition (SVD). 
In this approach, the channel matrix is diagonalized by taking SVD and removing the two unitary matrices through multiplication as a precoder and post-processing to the transmitter and receiver, respectively. 
Data streams are then sorted in decreasing order without creating any interference. 
To maximize system capacity, SVD is often combined with power allocation algorithms, e.g., water-filling (WF). WF implies allocating more power to the channels with higher signal-to-noise ratios (SNRs). 
 
There are other precoding techniques commonly used in multiuser MIMO systems. One example is ZF. ZF often involves channel inversion, using the pseudo-inverse of the channel matrix or other generalized inverses. Then, only diagonal terms remains, and the removed are off-diagonal terms, which are interference. As a result, this nulls the inter-user interference and achieves, when equal power is allocated, an equal received SNR for each data stream.

\subsection{SVC}

SVC is a useful technique to compress video content into a bit stream from which several decodable video streams can be extracted. It basically adopts a layered structure for compression, i.e., a bit stream is composed of a base layer representing the lowest quality version of the video data and one or more enhancement layers used to produce enhanced quality versions. Figure \ref{fig:SVC} illustrates three types of scalability: spatial scalability, quality (or SNR) scalability, and temporal scalability~\cite{jongseok12}. These refer, respectively, to the possibility of extracting video sequences having different spatial resolutions, different image quality levels, and different frame rates. Combinations of the three scalability options provide various versions of the video sequence with the corresponding bitrates and quality. 

Once video content has been encoded with SVC, the bit stream can, without the necessity of re-encoding, be used for efficient and adaptive transmission over error-prone networks. Suppose that the network condition from the server to the client is too poor to transmit the whole bit stream, or that the client's device is incapable of decoding and displaying the highest quality version of the content. In such cases, only part of the bit stream can be transmitted to the client to reduce the bitrate (and consequently quality) and meet the network or device constraint. Furthermore, if the network condition changes over time, the amount of the video data to be kept or discarded can be adaptively adjusted.

\subsection{Error control methods in cross-layer design}
In the physical layer, the likely data rate and error robustness of the system may be determined by using  modulation and coding (MC) scheme index values. If too many errors are being experienced, the MC value can be lowered thus reducing the error rate but at the cost of  slower data rate. An AMC scheme is performed to find the optimal modulation order and coding rate. 

In the application layer, erroneous packets need to be corrected for decoding. A simple way to do this is to replace the erroneous frame with a copy of the previous frame.
 
\section{Video transmission techniques in MIMO}
\label{MIMO}
Table \ref{tab:1} summarizes state-of-the-art video transmission algorithms for MIMO systems based on cross-layer design. Generally, according to the MIMO option one chooses, different error and transmission rates would be achieved. Unlike SVD-WF, which, by the MIMO option itself, determines power control, ZF beamforming (ZFBF) is able to independently control its power allocation. Thus, using ZFBF would guarantee a high degree of freedom and also yield advantages of being able to control the error rate of a channel. Given a MIMO option, whether to use ACS or adaptive modulation depends on the target system and trade-off issue. In the case of ACS, because high code rate means high redundancy, we will have some decrease in transmission rate. On the other hand, in the case of AM, error rate increases when modulation order gets higher and vice versa.

In~\cite{Song2008520}, the authors presented a quality of service (QoS)-aware SVC transmission framework in MIMO systems. In this framework, the authors made two assumptions. First, the data rate of the base layer should be maintained at all times to guarantee the minimum video QoS. Second, different protection levels are implemented between the enhancement layers, as they have different priorities. To maximize the QoS of SVC video, adaptive modulation and power allocation were adopted. This fully enhances, as a result, the MIMO capacity and allows different channel quality between the sub-channels. To guarantee the data rate of the base layer, the authors suggested to always assign the best quality sub-channel to the base layer and allocate power with fixed modulation. The rest of the power, which subtracts the assigned power from the total transmit power, is adaptively allocated to the enhancement layers. Adaptive modulation is only implemented for the enhancement layers.
Their optimization approach maximizes the total transmission rate by selecting power and modulation orders within the target bit error rate constraint. The power allocation step for the base and enhancement layers is repeated until the optimization approach bounds to the maximum data throughput based on a target bit error rate. They measured the performance of the framework in terms of the average peak signal-to-noise ratio (PSNR), which is a typical image quality metric.

\begin{table*}[t]
\centering
\caption{Summary of video transmission technologies with SVC in MIMO systems.(S: spatial scalability, Q: quality scalability, T:~temporal scalability, AM: adaptive modulation, RCPC: Rate-Compatible Punctured Convolution codes)}
\label{tab:1}       
\begin{center}
\begin{tabular}{p{0.4cm}p{2cm}p{1.5cm}p{1.5cm}p{1.5cm}p{1.3cm}p{5.1cm}}
\hline\hline\noalign{\smallskip}
Ref.& MIMO options& UEP& Channel coding&Scalability&Quality metric &Optimizations approaches \\ 
\hline\hline\noalign{\smallskip}
\cite{Song2008520} &SVD-WF (perfect CSI) &ACS, AM, Power~allocation &Reed-Solomon &S, Q, T&PSNR& Maximizing the total transmission rate by selecting power and modulation within the target bit error rate constraint\\ \hline

\cite{Zan201302}	&SVD,   Codebook (partial CSI)&AMC, Subcarrier~allocation &None&Q, T&PSNR& Maximizing the average SINR at the receivers and maximizing the transmission rate of each subcarrier by selecting MC within the SINR constraint\\\hline
\cite{Maodong201311}	&ZFBF&AM, Power~allocation &None&Q, T& PSNR& Maximizing the network efficiency by selecting bitrate, power, and modulation within the total transmit power constraint\\\hline
\cite{Wassim2013}&SVD-WF &ACS, AMC, Power~allocation &RCPC&S, Q, T&PSNR&Minimizing the total amount of distortions by selecting MC within the total transmit power constraint\\\hline
\cite{Chen2014}	&SVD&ACS, AM, Power~allocation &Reed-Solomon &Q, T&PSNR, SSIM& Maximizing the quality by finding optimal power per channel within the total transmit power constraint \\\hline
\noalign{\smallskip}\hline\noalign{\smallskip}

\end{tabular}
\end{center}
\end{table*}

\begin{table*}
\caption{Usage of UEP methods in introduced MIMO systems.}
\label{tab:2}
\begin{center}
\begin{tabular}{c|c|c|c|c|c}
\hline
&Adaptive &Adaptive &Adaptive channel   &   Power allocation or & MIMO configuration\\
&modulation (AM)&coding rate (AC) &selection (ACS)  &   subcarrier allocation& \\\hline\hline
\cite{Song2008520}	&		$\bigcirc$&	$\times$	&Not specified	&$\bigcirc$&$4\times4$ MIMO\\	
\cite{Zan201302}		&		$\bigcirc$&	$\bigcirc$	&Not specified		&$\bigcirc$ (subcarrier)&OFDM-MIMO   (multiuser)\\
\cite{Maodong201311}&		$\bigcirc$&	$\times$	&Not specified	&$\bigcirc$&OFDM-MIMO (multiuser-fairness)\\
\cite{Wassim2013}		&		$\bigcirc$&	$\bigcirc$	&$\bigcirc$		&$\bigcirc$&$4\times4$ MIMO\\
\cite{Chen2014}	&		$\bigcirc$&	$\times$	&$\bigcirc$	&$\bigcirc$&$4\times4$ MIMO\\\hline
\end{tabular}
\end{center}
\label{tab:2}
\end{table*}

In~\cite{Zan201302}, the authors also proposed a cross-layer framework for efficient broadcasting of scalable video over downlink MIMO orthogonal frequency division multiplexing (OFDM) systems. They considered the heterogeneity of both video sources and wireless channels of users. That is, they allocated resources not only between users but also between video layers using UEP. The goal was to maximize the average PSNR of all users while meeting the base layer's bandwidth constraint, where PSNR of a video stream is calculated using a rate-quality model.
For this, AMC, subcarrier allocation, and selection of SVD (with or without a codebook technique) were performed per user and per SVC video layer.

Another multiuser MIMO-OFDM framework was proposed in~\cite{Maodong201311}. These authors, unlike in~\cite{Zan201302}, used adaptive modulation and power allocation to maximize network efficiency and fairness between users. They employed ZFBF since it supports the close-optimal capacity and has a simple implementation. 
The authors also considered MAXMIN fairness for the base layer for every user. This way all users fairly receive their base layers. The authors then exploited a throughput-maximizing scheme for the enhancement layers. To reflect the visual experience for multiple users, a utility function was used, which takes into account the characteristics of different video content. The utility function was maximized by selecting an optimal power and modulation order of the enhancement layer sub-channels within the constraint of the total transmit power. It was shown that the framework yields better video quality in terms of average PSNR  than other frameworks such as round-robin, link gain-based resource allocation, multicarrier maximum sum rate, and so forth.

%
The authors in~\cite{Wassim2013} sought an optimal solution for SVC video transmission over MIMO channels.
 They developed a distortion model for SVC by considering loss of enhancement quality layers, drift error by loss of enhancement quality layers, and loss of the base quality layer. In the model, three physical layer parameters\textendash power allocation parameters (\emph{w}), ECC rates (\emph{r}), and the modulation order (\emph{m})\textendash are related to the total distortion in terms of the sum of the mean square error (MSE) over pixels of each video frame. A solution minimizing the total distortion is obtained by choosing optimal combinations (\emph{w, r, m}). The authors also developed a power allocation algorithm with rapid convergence using a fixed modulation order and ECC rate.


The authors in~\cite{Chen2014} laid out an approach for a cross-layer design for video delivery to optimize QoE of end users. The scheme jointly considers both transmission errors in the physical layer and video source coding characteristics in the application layer. To maximize the QoE, they defined a utility function that multiplies a SVC quality function and a frame correction rate function. 
A near-optimal solution determining parameters of ACS, adaptive modulation, and power allocation was achieved by decomposing the original optimization problem into several convex optimization sub-problems. The authors showed the near optimality of their proposed scheme, in terms of measured utilities, by comparing it with the exhaustively searched optimal solutions. Simulations demonstrated the effectiveness of their scheme in terms of PSNR and structural similarity (SSIM)~\cite{Bovik04ssim} that is a representative perceptual quality metric.


The aforementioned studies have been developed as a way to enhance the quality for end users. 
It was believed that maximizing transmission rate (capacity) is equivalent to maximizing the quality (e.g.,~\cite{Song2008520,Zan201302,Maodong201311}).
However, the discrepancy between them has been recognized and thus direct measures of video quality have been adopted recently (e.g., minimizing distortions~\cite{Wassim2013}, maximizing QoE~\cite{Chen2014}).
In addition, PSNR was frequently used as a measure of video quality, whereas more perceptually meaningful measures such as SSIM began to be used (e.g.,~\cite{Chen2014}).

 In summary, Table~\ref{tab:2} distinguishes clearly which UEP methods are used in the
introduced frameworks. Combination of multiple UEP methods might increase adaptability but it surely increases computational complexity. The service operator should have a choice among video delivery frameworks by considering not only the performance of the framework but also user scenarios, MIMO configurations, computational complexities, and so on.

As discussed above, it is mostly assumed that the base layer contains the most important information and thus must have the highest priority~\cite{Song2008520},~\cite{Zan201302},~\cite{Wassim2013},~\cite{Chen2014}.  
However, we wonder whether the SVC layer priorities still hold for massive MIMO systems. Studies have yet to be carried out on SVC video transmission over massive MIMO systems. 
The next section describes our initial studies on this issue.
We first discuss error behavior in massive MIMO systems, which differs from that in (non-massive) MIMO systems. Then, we examine the effectiveness of UEP solutions based on the SVC layer priorities in massive MIMO systems.

\section{Video transmission in massive MIMO}
\label{massiveMIMO}
\subsection{Error characteristics in massive MIMO sysetms}
\label{sub:mMIMO}
Massive MIMO systems have been proposed such that each BS is equipped with orders of magnitude more antennas, e.g.,~32, 64, 128 or more. More dramatic multiplexing or diversity gains are possible when the number of antennas at the BS ($N_t$) is significantly larger than the number of users ($K$), i.e., $K \ll N_t$. We consider that a massive MIMO system consists of $KN_r\times N_t$ channels with $K$ users and $N_r$ receive antennas for each user. With an increase in the number of antennas at the BS, linear precoders are shown to be near-optimal in terms of throughput~\cite{Caire2003}. Thus, for simplicity, ZF is used as a precoder. 

For massive MIMO systems, using the law of large numbers, the received SNR for each data stream of the $k$-th mobile station (MS) is expressed in terms of the number of antennas, transmit power, and number of users: i.e., $P_{k,i}(N_t-KN_r)$, while $P_{k,i}$ means transmit power in the $i$-th stream for the $k$-th MS~\cite{yglim}. 
Furthermore, the bit error probability ($\overline{P}_{b}$) in massive MIMO systems with a ZF precoder can be expressed as follows:
\begin{align}
 \label{fber}
 \nonumber
\overline{P}_{b} \approx \frac{1}{\log_2 M} \text{erfc} \Bigg\{\sqrt{\frac{P_{k,i}(N_t-KN_r)}{KN_r}}\sin\frac{\pi}{M}\Bigg\}
\end{align}
where $M$ is the modulation order. Thus, the error probability in massive MIMO systems is sensitive to changes with respect to power ($P_{k,i}$). 
  
This error probability can be transformed into packet error rate (PER) by merging the effect of the length of a SVC video packet: i.e., PER=$1-(1-\overline{P}_{b})^L$, where $L$ is the packet length~\cite{Goldsmith}. The size of each packet inherently varies depending on the amount of information in a video frame. Since the packet length has an exponential influence on PER, the PER rapidly increases as the packet length increases. 

\subsection{Perceived quality by using unequal power allocation}
\label{sub:UPA}
To examine the effectiveness of UEP in a massive MIMO system~\cite{soojin14}, we used six original video sequences that have different spatial and temporal content complexities. They were chosen from the SVT High Definition Multi Format Test Set~\cite{SVT} (\textit{CrowdRun,} and \textit{ParkJoy}) and the Live Video Quality Database~\cite{live} (\textit{pa1, mc1, sf1,} and \textit{sh1}). In our work, we encoded the base layer with a resolution of 176$\times$144 pixels at 15 fps, where the quantization parameter (QP) value was set to 28. For the enhancement layer we used a resolution of 352$\times$288 pixels at 30 fps and a QP value of 26. These two layers are received through separate receive antennas (i.e., $N_r$=2).

\begin{figure}[!tp]
	\begin{center}
		\begin{tabular}{cc}
			\includegraphics[width=1.0\linewidth]{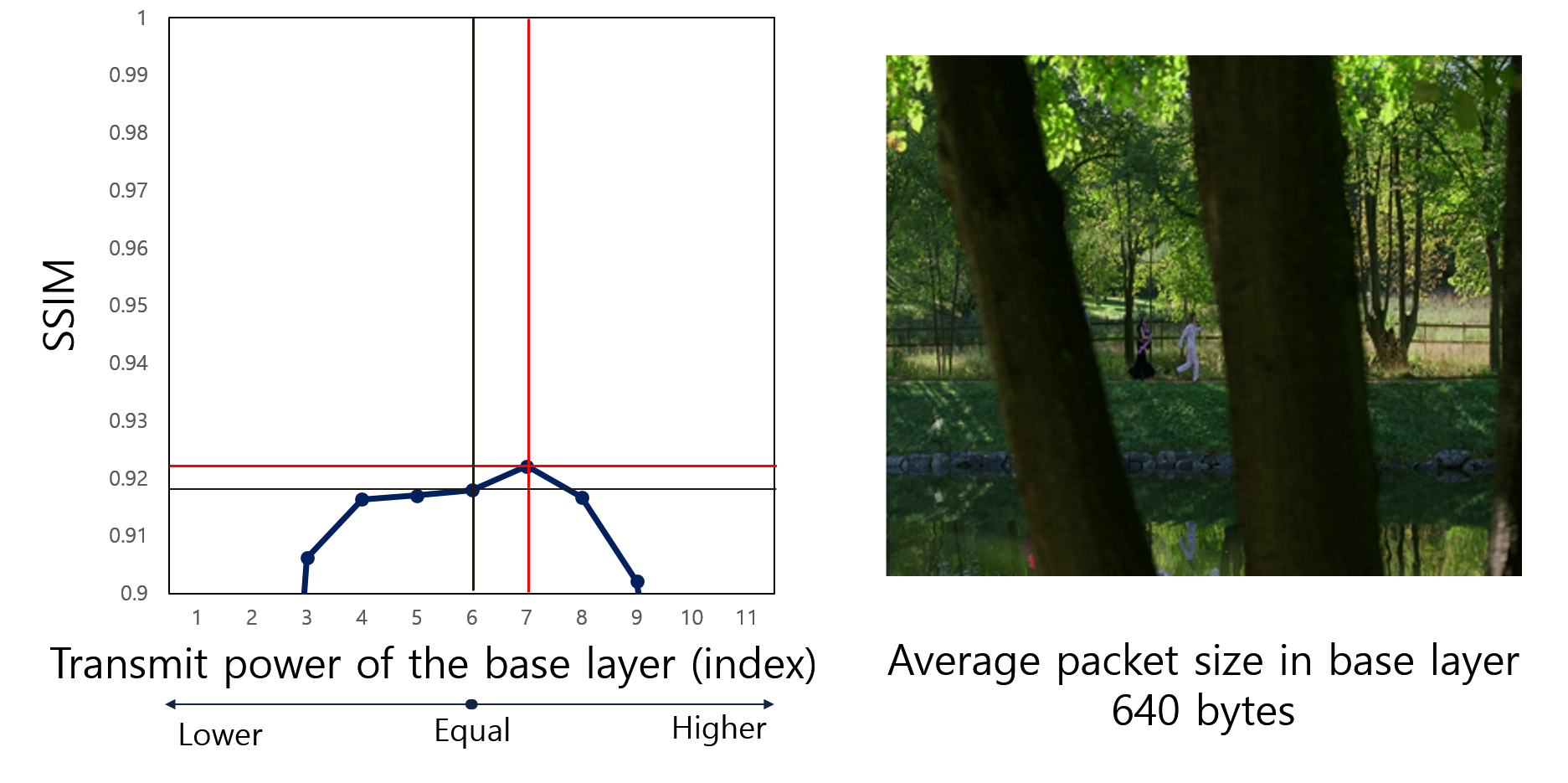}\\
			(a)\\\\
			\includegraphics[width=1.0\linewidth]{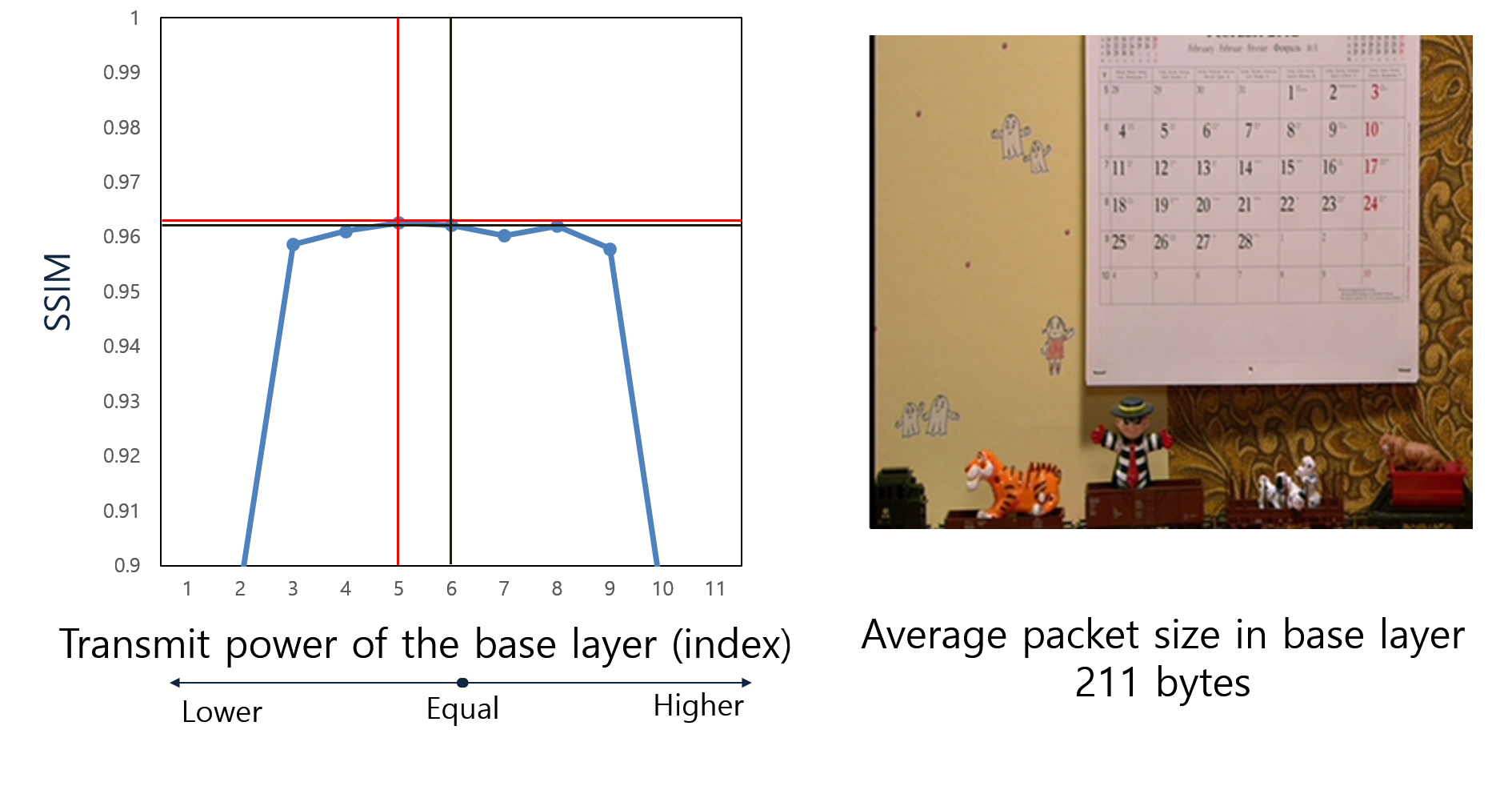}\\
			(b)\\\\
			\includegraphics[width=1.0\linewidth]{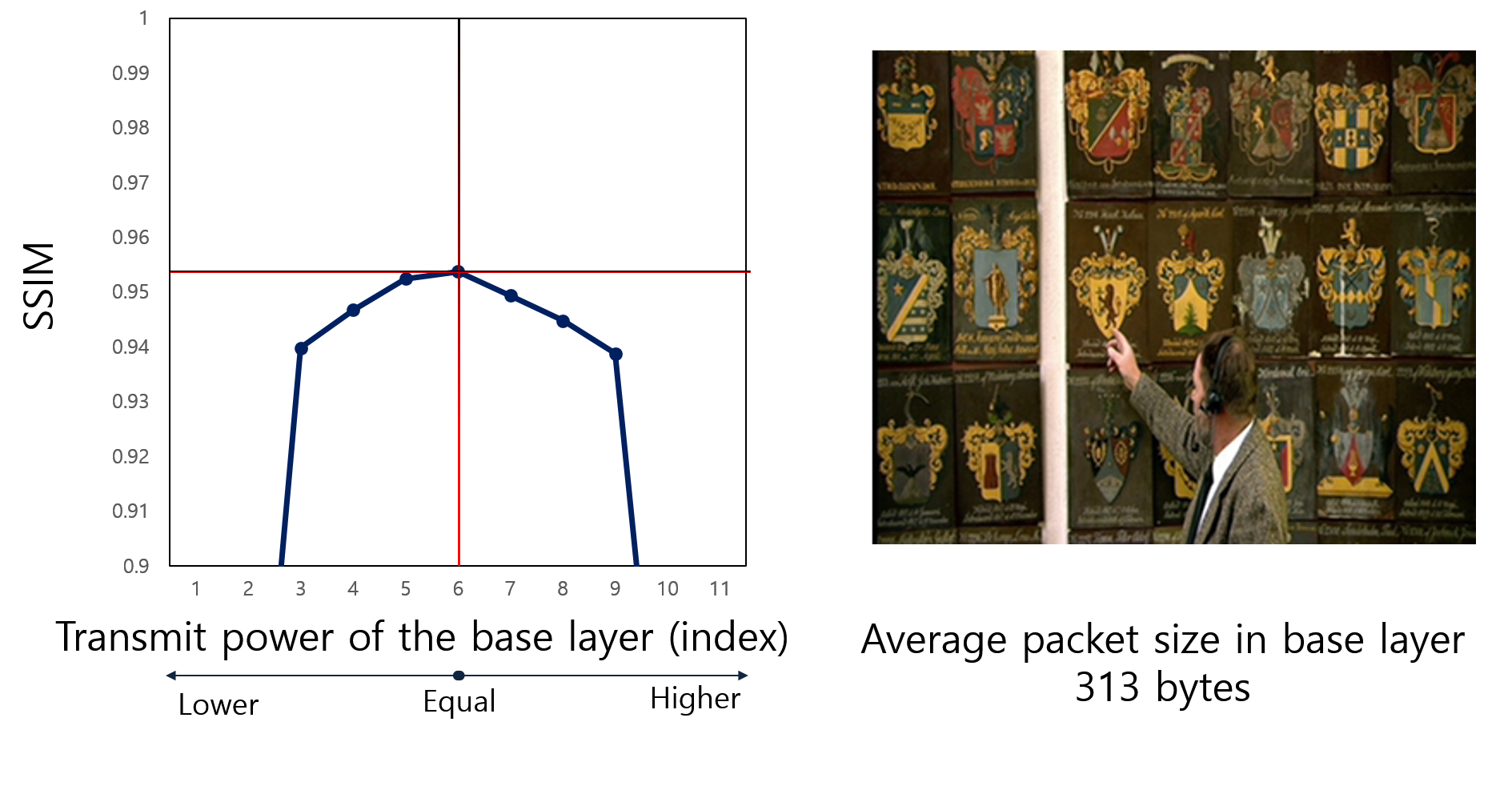}\\
			(c)\\
		\end{tabular}
	\end{center}
	\caption{SSIM vs. transmit power of the base layer when the total transmit power is 5.50~dB. The cross point of the black vertical and horizontal lines indicates equal power allocation ($2.48$dB) for the base layer and the enhancement layer. The cross point of the red vertical and horizontal lines indicates the case showing the highest SSIM value. A `lower' (or `higher') region means that the transmit power of the base layer is lower (or higher) than the transmit power of the enhancement layer. Thumbnails of the corresponding video sequences are also shown.  (a) \textit{ParkJoy} (b) \textit{mc1} (c) \textit{sh1}}
	\label{res:ssim}
\end{figure}
We applied ACS and power allocation and used ZF precoding.
In our simulation, we set $P_{k}=5.50$ dB, which corresponds to the total PER around 1\% when equal powers are allocated  (i.e., $P_{k,1}=P_{k,2}=2.48$ dB). The power range for UEP was set as 1.05 dB to 3.58 dB, which results in PER in the range of $1\%$ to $3\%$. This PER range can often be obtained in the typical wireless channels. We simulated several combinations of $P_{k,1}$ and $P_{k,2}$ in the chosen power range for giving unequal protection to the SVC layers.

\begin{figure*}[t]
	\centering
	\subfigure[]
	{\includegraphics[width=0.49\textwidth,keepaspectratio]{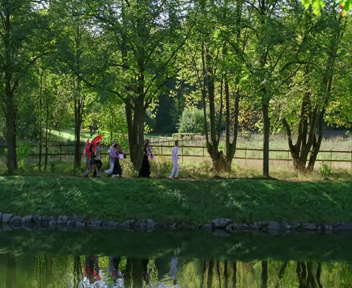}} 
		\subfigure[]
	{\includegraphics[width=0.49\textwidth,keepaspectratio]{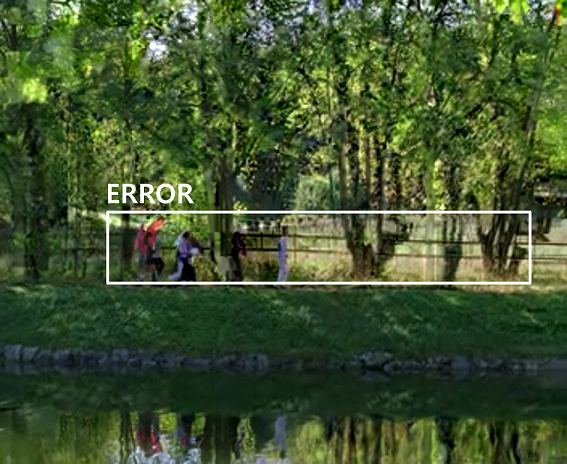}} %
	\caption{Example snapshots of the transmitted video, ParkJoy, (a) with the proposed algorithm (which showed the best quality) and (b) without the algorithm.}
	\label{Fig:thebest}	
\end{figure*}
Figure \ref{res:ssim} shows representative results of our simulation in terms of SSIM for measuring video quality. In Figure~\ref{res:ssim}(a), the maximum SSIM is obtained when the base layer power is higher than the other. In Figure \ref{res:ssim}(b), the maximum SSIM is obtained when a higher power is allocated to the enhancement layer than the base layer. In Figure \ref{res:ssim}(c), on the other hand, UEP never results in better quality than the equal power allocation. Among the three sequences, \textit{ParkJoy} and \textit{mc1} have, respectively, the largest and smallest average packet sizes of the base layer.
This implies that the sizes of packets of the base layer play an important role in determining the video quality with respect to the allocated power to each SVC layer.
 As shown in Section \ref{sub:mMIMO}, the massive MIMO system shows a large change in bit error probability even for a slight change in power. Furthermore, the effect of a bit error, in terms of PER, increases exponentially with respect to the packet size. 
 
Hence, it is true that the base layer is important, but this does not mean that the base layer always needs more power, as suggested by the previous work in MIMO systems, to maximize the overall quality. Thus, the effectiveness of a priority-based power allocation algorithm for SVC could, in practice, be interfered with by other factors such as network characteristics and content information.

Next, we attempt to model the relationship between the transmit power and the perceived quality based on the results in Fig. 3. We note that the curves in the figure can be modeled by quadratic functions. Then, the three curves can be considered as horizontal translations depending on the content characteristics. Therefore, we use the following function for regression:
\begin{align}
SSIM = a\cdot P_{k,1} ^2 + b\cdot P_{k,1} + c\cdot SI+ d \cdot TI + e
\end{align}
where, $P_k=P_{k,1}+P_{k,2}$ is the  total transmit power of the $k$-th user, and spatial information (SI) and temporal information (TI) are measures of content complexity in the spatial and temporal domains, respectively \cite{itup910}.

The five model parameters were determined as $a=-9.8301$, $b=-8.5383$, $c=0.3045$, $d=-0.0042$, and $e=15.3376$.
The performance of the regression model was measured as $0.92$ in terms of Pearson correlation coefficient. 
This model can be used to obtain the optimal power allocation between the two SVC layers for the given content in order to maximize the perceived quality of the user.
Figure~4 shows the snapshots of transmitted video, ParkJoy, when transmitted with unequal power allocation and with conventional equal power allocation. From this result, we can see that the proposed approach is applicable for finding the best quality for given $P_k$ by allocating different power per SVC layer (i.e., $P_{k,i}$). By demonstrating the quality difference between the video transmitted with the proposed algorithm and the other without the algorithm, we clarify the significance of the issue.\footnote{The full video comparison is available at http://www.cbchae.org/}. Note that this is our preliminary result for the regression model with only three contents. For our future work, we will advance our study with various parameters, network conditions, contents and so forth.

\section{Conclusion}
\label{Con}
In this article, we have investigated the problem of QoE-optimized SVC video transmission over massive MIMO systems. We first reviewed the state-of-the-art methods for SVC transmission in MIMO, which mostly adopted UEP based on SVC layer priority. However, we argued that such approaches may not be optimal for massive MIMO systems.
Although massive MIMO channels provide a dramatic increase in spectral efficiency and error robustness, with only a small change in the amount of transmit power, PER changes drastically. The massive MIMO channel error characteristic and the amount of information of content were combined to yield the formula of PER. Our experimental results suggest that priority between SVC layers does exist, but the highest priority must not necessarily be given to the base layer. Depending on the content characteristics, the base or enhancement layer has the priority, or both layers have the same priority. Unlike MIMO systems, both system and content characteristics impact the effectiveness of UEP solutions. 

This work though represents an initial study into SVC transmission in massive MIMO systems, providing guidelines for developing cross-layer video transmission frameworks in massive MIMO systems. 
Starting with this result, we need to find optimal resource allocation solutions for SVC transmission in massive MIMO systems. It is also necessary to explore various scenarios. In this article, for example, we have only considered video transmission based on broadcasting. If the BS, however, has to support heterogeneous devices or is multicasting, where data are sent from multiple sources to multiple destinations, we should consider UEP methods for multi-user or multi-content scenarios while SVC layer priority per user are considered simultaneously. To elevate the video transmission in massive MIMO systems, we will more thoroughly investigate the relationship between error probability and QoE in our future work. 

\bibliographystyle{IEEEtran}

\bibliography{reference}
\balance

\renewenvironment{IEEEbiography}[1]
{\IEEEbiographynophoto{#1}}
{\endIEEEbiographynophoto}

\begin{IEEEbiography}{Soo-Jin Kim} received her BS degree in computer engineering from Kwangwoon University in Korea at 2005. Now, she is integrated master and doctoral student in the School of Integrated Technology, Yonsei University in Korea. Before joining Yonsei, she was with LG electronics and KT for 7.5 years as a research engineer from 2005 to 2012. Her research interests include the cross-layer optimizations between the channel coding and source coding for multimedia delivery in mobile communications. 
\end{IEEEbiography}

\begin{IEEEbiography}{Gee-Yong Suk} received his BS degree in electrical and electronics engineering from Yonsei University in Korea at 2016. Now, he is a graduate student in the School of Integrated Technology, Yonsei University in Korea. His research interests include millimeter wave communications, MIMO communications, 5G networks, molecular communications, and estimation theory.
\end{IEEEbiography}

\begin{IEEEbiography}{Jong-Seok Lee}(SM'14) received his Ph.D. degree in electrical engineering and computer science from the Korea Advanced Institute of Science and Technology, Korea. He was a Research Scientist with the Swiss Federal Institute of Technology in Lausanne (EPFL), Switzerland. He is currently an Associate Professor with the School of Integrated Technology, Yonsei University, Korea. He has authored or coauthored more than 100 publications. His research interests include multimedia signal processing and machine learning. He currently serves as an Editor for IEEE Communications Magazine and Signal Processing: Image Communication.
\end{IEEEbiography}

\begin{IEEEbiography}{Chan-Byoung Chae}(SM'12) is an associate professor in the School of Integrated Technology, Yonsei University. Before joining Yonsei University, he was with Bell Labs, Alcatel-Lucent, Murray Hill, New Jersey, as a member of technical staff, and Harvard University, Cambridge, Massachusetts, as a postdoctoral research fellow. He received his Ph.D. degree in electrical and computer engineering from the University of Texas at Austin in 2008. He was the Best Young Professor from Yonsei University (2015), the recipient/co-recipient of the IEEE INFOCOM Best Demo Award (2015), the IEIE/IEEE Joint Award for Young IT Engineer of the Year (2014), the KICS Haedong Young Scholar Award (2013), the IEEE Signal Processing Magazine Best Paper Award (2013), the IEEE ComSoc AP Outstanding Young Researcher Award (2012), the IEEE Dan. E. Noble Fellowship Award (2008), and two Gold Prizes (1st) in the 14th/19th Humantech Paper Contest. He currently serves as an Editor for IEEE Transactions on Wireless Communications, the IEEE/KICS Journal on Communications Networks, and IEEE Transactions on Molecular, Biological, and Multi-scale Communications.
\end{IEEEbiography}

\end{document}